\documentclass[prd,twocolumn,superscriptaddress,altaffilletter,nofootinbib]{revtex4}

\usepackage{amssymb}
\usepackage{amsfonts}
\usepackage{amsthm}
\usepackage{graphicx}
\usepackage{graphicx,color}
\usepackage{amsmath}
\usepackage{latexsym}

\newtheorem{theorem}{Theorem}
\newtheorem{corollary}{Corollary}[theorem]

\newcommand{\be}{\begin{equation}}
\newcommand{\ee}{\end{equation}}
\newcommand{\bea}{\begin{eqnarray}}
\newcommand{\eea}{\end{eqnarray}}
\newcommand{\der}{\partial}

\newcommand{\bet}{\begin{theorem}}
\newcommand{\eet}{\end{theorem}}
\newcommand{\bec}{\begin{corollary}}
\newcommand{\eec}{\end{corollary}}

\begin{document}



\title{Revisiting purely kinetic k-essence}



\author{Israel Quiros}\email{iquiros@fisica.ugto.mx}\affiliation{Departamento Ingenier\'ia Civil, Divisi\'on de Ingenier\'ia, Universidad de Guanajuato, C.P. 36000, Gto., M\'exico.}

\author{Tame Gonzalez}\email{tamegc@ugto.mx}\affiliation{Departamento Ingenier\'ia Civil, Divisi\'on de Ingenier\'ia, Universidad de Guanajuato, C.P. 36000, Gto., M\'exico.}

\author{Ulises Nucamendi}\email{ulises.nucamendi@umich.mx}\affiliation{Facultad de Ciencias Físico-Matemáticas, Universidad Michoacana de San Nicolás de Hidalgo, Edificio Alfa, Ciudad Universitaria, 58040 Morelia, Michoacán, Mexico.}

\author{Roberto De Arcia}\email{rc.dearcia@ugto.mx}\affiliation{Facultad de Ciencias Físico-Matemáticas, Universidad Michoacana de San Nicolás de Hidalgo, Edificio Alfa, Ciudad Universitaria, 58040 Morelia, Michoacán, Mexico.}

\author{Francisco Antonio Horta Rangel}\email{anthort@ugto.mx}\affiliation{Departamento Ingenier\'ia Civil, Divisi\'on de Ingenier\'ia, Universidad de Guanajuato, C.P. 36000, Gto., M\'exico.}



\begin{abstract}
In this paper, we perform a dynamical systems study of the purely kinetic k-essence. Although these models have been studied in the past, a full study of the dynamics in the phase space incorporating the stability conditions for theoretical consistency is lacking. Our results confirm in a very rigorous and clear way that these models i) can not explain in a unified way the dark matter and dark energy components of the cosmic fluid and ii) are not adequate to explain the existing observational evidence, in particular the observed amount of cosmic structure. 
\end{abstract}



\maketitle


\section{Introduction}
\label{intro} 


The k-essence was introduced with the intention of explaining the early inflation \cite{picon_damour_plb_1999, garriga_plb_1999}. Eventually, it became clear that alternatively it may explain the present stage of accelerated expansion of the universe \cite{chiba_prd_2000, picon_mukhanov_prl_2000, picon_mukhanov_prd_2001, chime_mpla_2004, chime_prd_2004, bose_prd_2009, scherrer_prl_2004}. In general, the k-essence models are given by the following action:

\begin{align} S=\int d^4x\sqrt{-g}\left[\frac{M^2_\text{pl}}{2}\,R+K(\phi,X)+{\cal L}_m\right],\label{action}\end{align} where $R$ is the curvature scalar, $K=K(\phi,X)$ is an arbitrary function of the scalar field and its kinetic energy density $X:=-(\der\phi)^2/2$ and ${\cal L}_m={\cal L}_m(\chi,\der\chi,g_{\mu\nu})$ is the Lagrangian of the matter fields, collectively denoted by $\chi$. Here we use the following shorthand notation: $(\der\phi)^2\equiv g^{\mu\nu}\der_\mu\phi\der_\nu\phi$, $K_\phi\equiv\der K/\der\phi,$ $K_X\equiv\der K/\der X,$ $K_{XX}\equiv\der^2K/\der X^2,$ etc. Our metric signature is $(-+++)$. In what follows we use the units system where the Planck mass $M_\text{pl}=1$. 

Since we are looking for the cosmological dynamics of k-essence, we assume the Friedmann-Robertson-Walker (FRW) metric with flat spatial sections: $ds^2=-dt^2+a^2(t)\delta_{ik}dx^idx^k,$ where $a=a(t)$ is the scale factor and $t$ is the cosmological time. We shall focus on the purely kinetic k-essence, so we assume that the effective pressure $p_\text{eff}=K(\phi,X)=K(X),$ is an explicit function of the kinetic energy density only.

The equations of motion (EOM) that are derived by varying the action \eqref{action}, read:

\begin{align} 3H^2&=2XK_X-K+\rho_m,\nonumber\\
-2\dot H&=2XK_X+(w_m+1)\rho_m,\label{eom}\end{align} plus the conservation equation,

\begin{align} \dot\rho_\text{eff}=-3H(\rho_\text{eff}+p_\text{eff})=-6HXK_X,\label{k-cons}\end{align} where

\begin{align} \rho_\text{eff}=2XK_X-K,\label{rho-k}\end{align} is the effective energy density of the k-essence. Equation \eqref{k-cons} can be written also in the following more compact way,

\begin{align} \dot X=-6c^2_sHX.\label{kg-eom}\end{align} In the above equations the dots account for derivatives with respect to the cosmic time, $H\equiv\dot a/a$ and we are assuming a background matter fluid with energy density $\rho_m$ and equation of state (EOS) parameter $w_m$. Besides,

\begin{align} c^2_s=\frac{K_X}{K_X+2XK_{XX}},\label{sss}\end{align} is the squared speed of sound (SSS). 

The Friedmann equation (first equation in \eqref{eom}) can be written in the form of the following constraint:

\begin{align} \Omega_m=1-\Omega_\text{eff},\label{om}\end{align} where we have defined the dimensionless energy density parameters $\Omega_m\equiv\rho_m/3H^2$ and

\begin{align} \Omega_\text{eff}=\frac{\rho_\text{eff}}{3H^2}=\frac{2XK_X}{3H^2}-\frac{K}{3H^2}.\label{ok}\end{align} For the k-essence EOS parameter we obtain that,

\begin{align} w_\text{eff}=\frac{K}{2XK_X-K}.\label{wk}\end{align} For convenience we also rewrite the remaining EOM \eqref{eom}, \eqref{kg-eom} as it follows:

\begin{align} \frac{\dot H}{H^2}&=-\frac{XK_X}{H^2}-\frac{3}{2}(w_m+1)\Omega_m,\label{hdot}\\
\frac{\dot X}{HX}&=-6c^2_s.\label{xdot}\end{align}

The stability of the perturbations of \eqref{action} around the background requires that the following conditions were satisfied:

\begin{align} K_X+2XK_{XX}\geq 0,\;c^2_s\geq 0.\label{stab}\end{align} The left-hand side (LHS) bound is the so called ``no-ghost'' condition \cite{defelice_2012}. It amounts to requiring the perturbed Hamiltonian about a background solution to be positive so that it is associated with quantum stability \cite{fang_2014}. The right-hand side (RHS) condition in \eqref{stab} is required in order to avoid the Laplacian instability which is a classical and so, catastrophic instability.


Purely kinetic k-essence \cite{scherrer_prl_2004} was proposed as a unified model of dark matter (DM) and dark energy (DE). It is assumed that the effective pressure $p_\text{eff}=K$ is an extremum in some $X_0$, so $K_X(X_0)=0$, $K_{XX}(X_0)\neq 0$. From the EOM \eqref{eom} it then follows that, as the energy density of matter dilutes with the expansion, the spacetime tends to de Sitter space: $H\rightarrow H_0$, where 

\begin{align} H_0=\sqrt\frac{-K(X_0)}{3}.\label{de-sitter}\end{align} As seen it is required that at the extremum $K(X_0)<0$.\footnote{It should be said that the effective energy density of the k-essence $\rho_\text{eff}=2XK_X-K$ can be an extremum as well, if the condition $$\frac{d\rho_\text{eff}}{dX}=K_X+2XK_{XX}=0,$$ is fulfilled, i. e., if the no-ghost bound \eqref{stab} is saturated\label{foot1}.}

If we expand the SSS \eqref{sss} around the extremum: $X=X_0+\varepsilon$ ($\varepsilon$ is a small perturbation), since $K_X=K_{XX}\,\varepsilon,$ then

\begin{align} c^2_s\approx\frac{\varepsilon}{2X_0}.\nonumber\end{align} If substitute this back into \eqref{xdot} one obtains that $\dot\varepsilon\approx-3H_0\,\varepsilon$ $\Rightarrow$ $\varepsilon(t)=\varepsilon_0\,\exp{(-3H_0\,t)}$, so that perturbation decays with the expansion and the extremum is a stable solution. Notice, from equations \eqref{sss} and \eqref{wk}, that at the extremum $c^2_s=0$ while $w_\text{eff}(X_0)=-1.$ Thus, this solution corresponds to an EOS equivalent to a cosmological constant with zero effective sound speed \cite{scherrer_prl_2004}. 

The purely kinetic k-essence has other very interesting properties. As shown in \cite{scherrer_prl_2004} the k-essence fluid behaves like a low sound-speed fluid with a density that evolves like the sum of a DM component ($\rho_{DM}\propto a^{-3}$) and a DE component ($\rho_{DE}=$ const.) This is why the model is supposed to provide a unified description of dark matter and dark energy. This k-essence model differs from the standard $\Lambda$CDM model in that the dark-energy component has $c^2_s\ll 1$. This suppresses the integrated Sachs-Wolfe effect at large angular scales and, at the same time, enhances the matter power spectrum that dark energy clustering
induces at large scales \cite{scherrer_prl_2004, erickson_2002, dedeo_2004, fang_2014}.

In this paper, we apply the tools of the dynamical system theory in order to extract useful information about the generic solutions of the purely kinetic k-essence model, without the need to solve the EOM \eqref{eom}, \eqref{k-cons}. In this case, we trade the original EOM of the model by an autonomous system of ordinary differential equations (ASODE) in terms of the variables of some equivalent phase space. The critical points of this phase space (past and future attractors as well as saddle points) correspond to the asymptotic behavior of the dynamical system, i. e., to the exact solutions of the original EOM which are ``preferred'' by the dynamical system in the long term. 

Although there are dynamical systems studies of kinetic k-essence models in the bibliography \cite{yang_2011, fang_2014, chakra_2019}, the physical consequences drawn from these studies are complex and nonconclusive. For example, in none of these studies the stability conditions \eqref{stab} were considered (proper consideration of the stability conditions bounds the physical phase space to a narrow region of the whole mathematical space). There are several critical points that lie outside of the physical phase space and, consequently, should not be considered. In some of the studies \cite{yang_2011, fang_2014} the model $K(\phi,X)=V(\phi)(-X+X^2)$ is considered, so that there are no free parameters in the kinetic term to readjust the model. In addition, the choice of phase space variables is not the most adequate in some cases \cite{fang_2014, chakra_2019} as the chosen variables are neither normalized nor bounded. Besides, in none of the existing studies the phase space is rigorously defined, so that the exercise of finding critical points is sometimes futile. These features are required in order for the results to be independent of the physical units and for the whole phase space to be accessible, so that no critical points at infinity are lost. Take, for instance, the study of the model $K(\phi,X)=F(X)-V(\phi)$ in \cite{cervantes_prd_2013}. This work is an example of the required study of critical points at infinity when one coordinate chart is not enough to cover the whole phase space. 

In the present study, we consider all the above facts: 1) the stability conditions \eqref{stab}, 2) the normalized and bounded phase-space variables, and 3) the rigorous definition of the phase space. This helps us to draw clear physical conclusions from the study. We shall show, in particular, that there are no critical points that could be associated with dark-matter domination, which is expected to be correlated with saddle critical points (these are associated with transient behavior). This stage is required for the amount of observed cosmic structure to form. Consequently, for the types of kinetic k-essence models investigated in this paper, these are unable to provide a unified description of DM and DE, as claimed in \cite{scherrer_prl_2004}. Moreover, these models cannot fit existing data on structure formation.


The paper is organized as follows. In sections \ref{sect-mod-1} and \ref{sect-mod-2}, on the basis of the dynamical systems study, simple and generic models of kinetic k-essence are investigated, which are second-order polynomials in $X$ and in $\dot\phi$, respectively. These models reflect quite well the dynamics anticipated in \cite{scherrer_prl_2004}. Models of power-law type are briefly investigated in section \ref{sect-other}. Discussion of the results of this paper and brief conclusions are given in Section \ref{sect-discuss}.


\section{Model $K=-\alpha X+\beta X^2$}
\label{sect-mod-1}


In this section, we study the type of polynomial k-essence of \cite{chiba_prd_2000, yang_2011, fang_2014}:

\begin{align} K(X)=-\alpha X+\beta X^2,\label{pwl-mod}\end{align} where $\alpha$ and $\beta$ are nonnegative constant parameters ($\alpha$ is dimensionless, while $\beta$ has dimensions of $H^{-2}$). Notice that in \cite{chiba_prd_2000} and in many works after it, a new dimensionless variable is defined $X_\text{new}=\beta X/|\alpha|$, so that \eqref{pwl-mod} is written in the following way

\begin{align} K(X)=-X_\text{new}+X^2_\text{new}.\nonumber\end{align} Here we prefer to work with \eqref{pwl-mod} so that $\alpha$ and $\beta$ are adjustable parameters. 

The EOS parameter and the SSS for this model are given by

\begin{align} w_\text{eff}&=\frac{\beta X-\alpha}{3\beta X-\alpha},\nonumber\\
c^2_s&=\frac{2\beta X-\alpha}{6\beta X-\alpha},\label{par-mod}\end{align} respectively. The effective pressure $K=K(X)$ is a minimum at $X_0=\alpha/2\beta,$ where $K_{XX}(X_0)=2\beta>0$, while $K(X_0)=-\alpha^2/4\beta$ is a negative quantity as required by \eqref{de-sitter}: $H_0=\sqrt{\alpha^2/12\beta}$. In this case, the no-ghost condition \eqref{stab} reads $X\geq\alpha/6\beta,$, while the Laplacian stability requirement: $c^2_s\geq 0$, yields either $X\geq\alpha/2\beta,$ or $X<\alpha/6\beta$. As seen, both stability conditions in \eqref{stab} are jointly met only if $X\geq\alpha/2\beta$. 

The model \eqref{pwl-mod} meets all the features required by the analysis of \cite{scherrer_prl_2004} in the general case. It also represents a nice example to illustrate the existence of an extremum of the energy density of the k-essence \eqref{rho-k} (see footnote \ref{foot1}). In this case $\rho_\text{eff}$ is an extremum at $X_*=\alpha/6\beta$. It is a minimum since 

\begin{align} \left.\frac{d^2\rho_\text{eff}}{dX^2}\right|_{X_*}=6\beta>0.\nonumber\end{align} At the minimum the energy density of the k-essence is negative $\rho^*_\text{eff}=\rho_\text{eff}(X_*)=-\alpha^2/12\beta$, so that this is a unstable state. In the left and middle panels of FIG. \ref{fig1} this state is represented by the dashed curve. It is seen how the phase plane orbits diverge from this curve to the region below the curve and to the region above the curve, respectively. Orbits that evolve in the region below the curve are classically stable since $c^2_s\geq 0$, however these may develop ghosts which are relevant when quantum effects become important.


There are other alternative variants of this model. For instance, if assume $K(X)=\alpha X-\beta X^2$, the scalar field's effective pressure $K=K(X)$ is a maximum at $X_0=\alpha/2\beta,$ since $K_X(X_0)=0$ and $K_{XX}(X_0)=-2\beta<0.$ This case is not physical since at the maximum $K(X_0)=\alpha^2/4\beta>0$ is a positive constant so that from \eqref{de-sitter} it follows the unphysical result that $H_0^2=-\alpha^2/12\beta<0.$ The remaining possibility: $K(X)=\alpha X+\beta X^2,$ is not of cosmological interest since the effective pressure $K=K(X)$ does not have extremum and, besides, it can not produce accelerated expansion.


\begin{table*}\centering
\begin{tabular}{||c||c|c|c|c|c|c|c|c||}
\hline\hline
Critical Point &  $x$   &   $y$   & Existence & Stability  & $\Omega_m$ &  $w_\text{eff}$  & $q$ & $c^2_s$ \\
\hline\hline
$P_\text{mat}$     &  0   &  0   &  always & stable (isolated attractor) &  1 & 1  &  1/2  & 1 \\
\hline
$P_\text{bb}$     & 0   &   1   &   always & unstable (source point) &  1 & undef. & 1/2 & undef. \\
\hline
$P_\text{K}$     & 1  &   0   &   always & stable (isolated attractor) &   undef.  & undef. &  undef. & undef. \\
\hline
$P_\text{rad}$  &  1   &  1  &  always & unstable (saddle point) &  undef. & 1/3 & undef. & 1/3 \\
\hline
$P_\text{dS}$  &  $\frac{6}{7}$  &  $\frac{1}{13}$ & always & stable (isolated attractor) &  0 & -1 & -1 & 0 \\
\hline\hline
\end{tabular}\caption{Critical points of the dynamical system \eqref{dsyst} and their main properties. Several of these points and their stability properties can be found analytically but others, like $P_\text{bb}$ and $P_\text{k}$, have to be investigated numerically.}\label{tab1}\end{table*}



\subsection{Dynamical system}
\label{subsect-ds}


Sometimes, when the task of finding analytic solutions of cosmological equations is complex, it is more appropriate to trade the original EOM by an ASODE (dynamical system). This means that we move on to some abstract phase space that is spanned by a set of phase-space variables. We then apply the tools of the dynamical system theory \cite{wands_prd_1998, copeland_rev_2006, quiros_ejp_2015, quiros_rev_2019}. Critical (also singular) points of the ASODE in the phase space amount to asymptotic states to which solutions of the original EOM asymptote, either into the past or into the future. This picture is complemented with saddle equilibrium configurations which are not only unstable but metastable: phase-space orbits are attracted to this point in one direction and repelled in another direction. In fact, each critical point in the phase space corresponds to a generic or distinctive solution of the cosmological EOM. For a brief introduction to the application of the dynamical system's tools in cosmology and for specifications about the notation we use here, we recommend \cite{quiros_ejp_2015}.


An appropriate phase space in which to look for equilibrium configurations of the EOM \eqref{eom}, \eqref{kg-eom}, is spanned by the following dimensionless and bounded variables:

\begin{align} x=\frac{\alpha X}{\alpha X+H^2},\;y=\frac{\beta H^2}{\beta H^2+\alpha^2}.\label{xy-var}\end{align} We have that

\begin{align} \frac{\alpha X}{H^2}=\frac{x}{1-x},\;\beta H^2=\frac{\alpha^2y}{1-y}.\nonumber\end{align} Besides,

\begin{align} \frac{K}{H^2}&=-\frac{x(1-x-y)}{(1-x)^2(1-y)},\nonumber\\
\frac{XK_X}{H^2}&=-\frac{x(1-x-y-xy)}{(1-x)^2(1-y)},\nonumber\\
K_X&=-\frac{\alpha(1-x-y-xy)}{(1-x)(1-y)},\nonumber\\
XK_{XX}&=\frac{2\alpha\,xy}{(1-x)(1-y)},\label{usef}\end{align} are useful expressions as well.

In terms of the phase-space variables \eqref{xy-var} the effective EOS of the scalar field and the SSS reads,

\begin{align} w_\text{eff}&=\frac{1-x-y}{1-x-y-2xy},\nonumber\\
c^2_s&=\frac{1-x-y-xy}{1-x-y-5xy},\label{xy-sss}\end{align} respectively. Defining the following curves in the phase plane:

\begin{align} y_\bullet=\frac{1-x}{1+5x},\;y_*=\frac{1-x}{1+x},\label{curves}\end{align} where $y_\bullet$ represents the loci where the energy density of k-essence $\rho_\text{eff}$ \eqref{rho-k} is an extremum, the stability conditions \eqref{stab} amount to:

\begin{align} K_X+2XK_{XX}\geq 0\;\Rightarrow\;y\geq y_\bullet,\nonumber\\
c^2_s\geq 0\;\Rightarrow\;y<y_\bullet\;\text{or}\;y\geq y_*.\label{xy-stab}\end{align} The stability conditions are jointly met at $y\geq y_*$.

The EOM \eqref{om}, \eqref{hdot} and \eqref{xdot}, can be written in terms of the variables $x$, $y$ as follows:

\begin{align} \Omega_m&=1+\frac{x(1-x-y-2xy)}{3(1-x)^2(1-y)},\label{xy-om}\\
\frac{\dot H}{H^2}&=\frac{x(1-x-y-xy)}{(1-x)^2(1-y)}-\frac{3}{2}(w_m+1)\Omega_m,\label{xy-hdot}\\
\frac{\dot X}{HX}&=-6\frac{1-x-y-xy}{1-x-y-5xy}.\label{xy-xdot}\end{align} The physical requirement that the normalized energy density of matter must be non-negative: $\Omega_m\geq 0$, amounts to

\begin{align} y\leq y_\dag=\frac{3-5x+2x^2}{3-5x+5x^2}.\label{y-dag}\end{align} 

The original EOM \eqref{eom} is traded by the following two-dimensional dynamical system:

\begin{align} \frac{dx}{dN}&=x(1-x)\left(\frac{\dot X}{HX}-2\frac{\dot H}{H^2}\right),\nonumber\\
\frac{dy}{dN}&=2y(1-y)\frac{\dot H}{H^2},\nonumber\end{align} where $N=\ln a$. In order for this ASODE to be defined everywhere, it is recommended to introduce the new time-ordering variable: $d\tau=dN/(1-x)^2(1-y)$. We get

\begin{align} x'&=x(1-x)^3(1-y)\left(\frac{\dot X}{HX}-2\frac{\dot H}{H^2}\right),\nonumber\\
y'&=2y(1-x)^2(1-y)^2\frac{\dot H}{H^2},\label{dsyst}\end{align} where a prime denotes the derivative with respect to $\tau$. 

\begin{figure*}[tbh]\centering
\includegraphics[width=5.5cm]{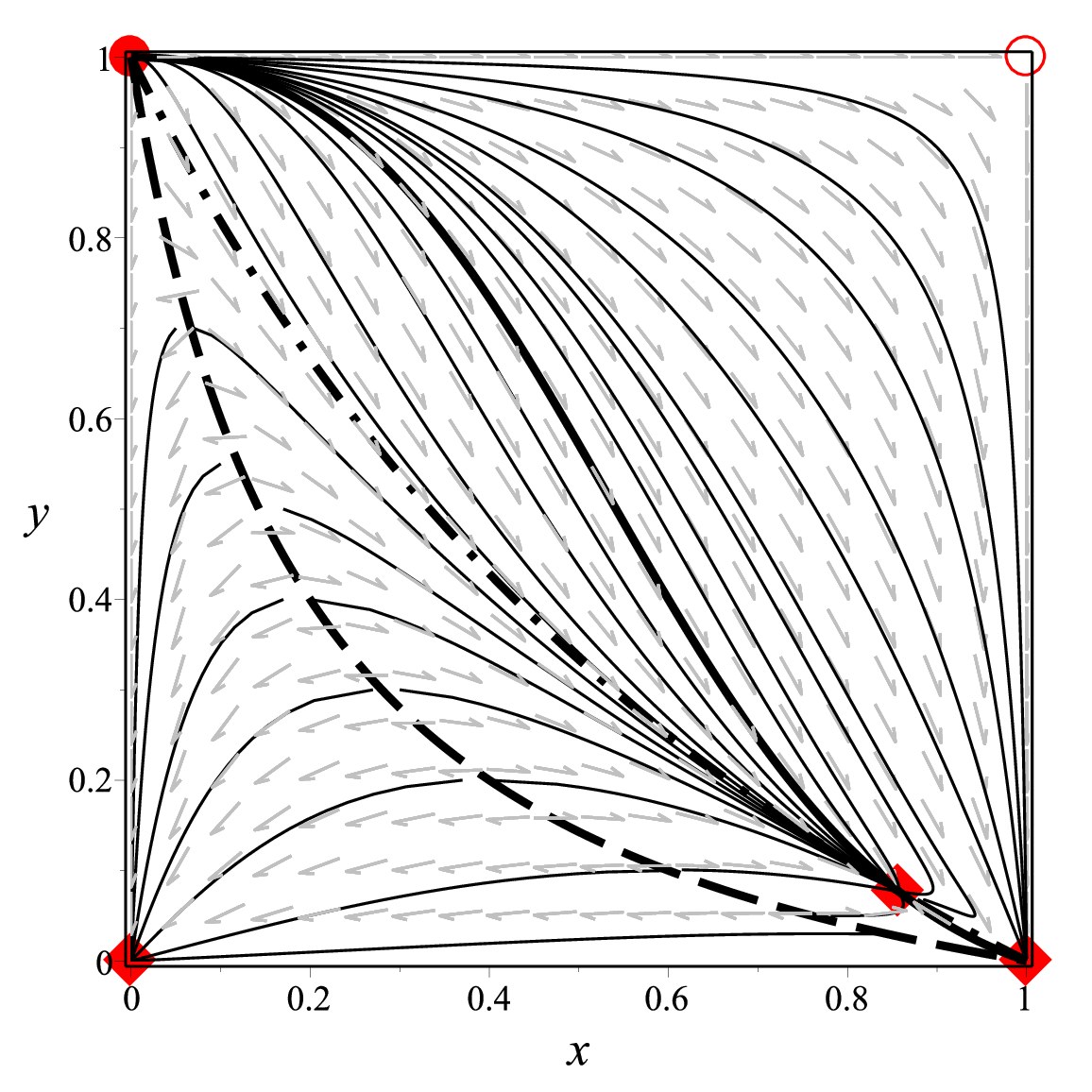}
\includegraphics[width=5.5cm]{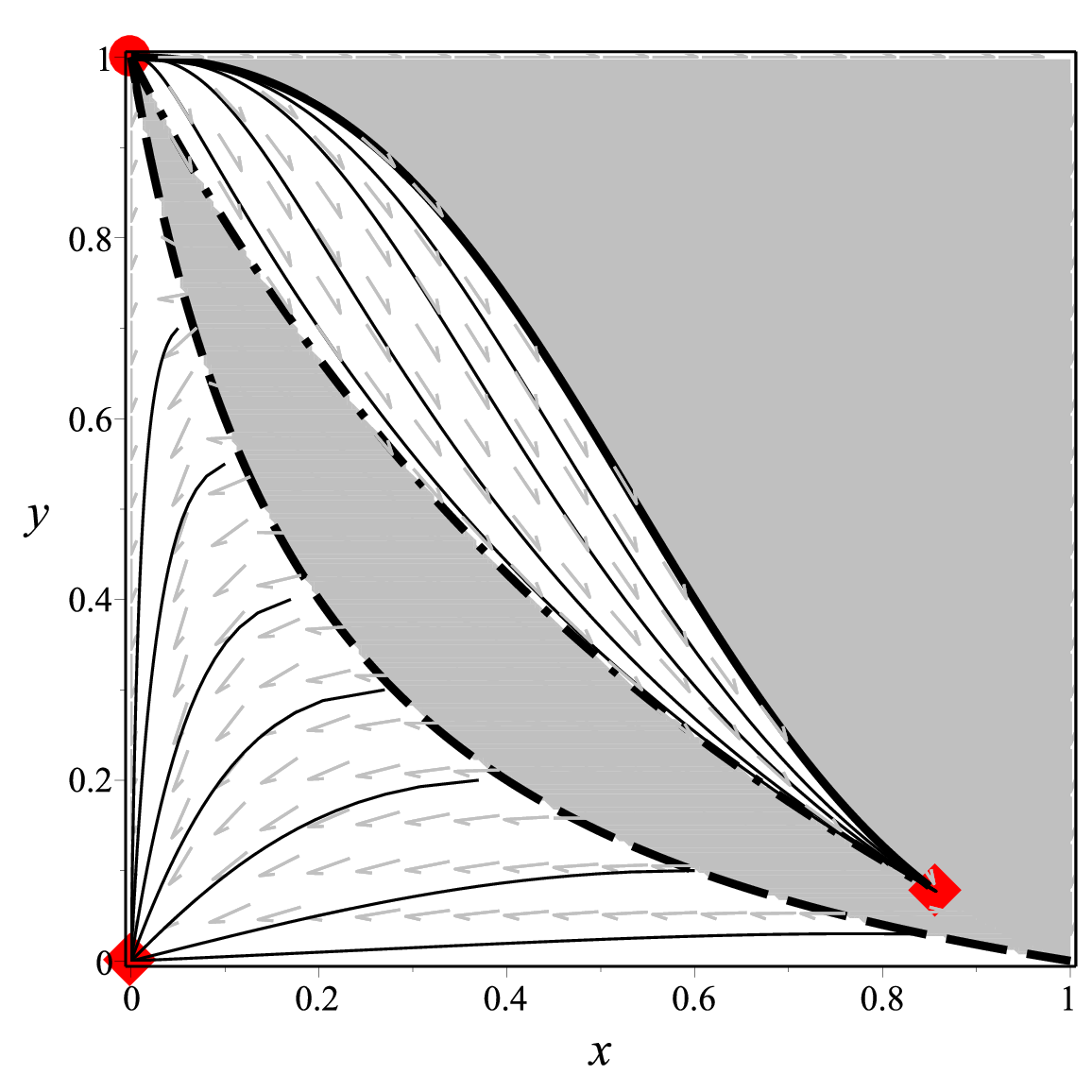}
\includegraphics[width=5.5cm]{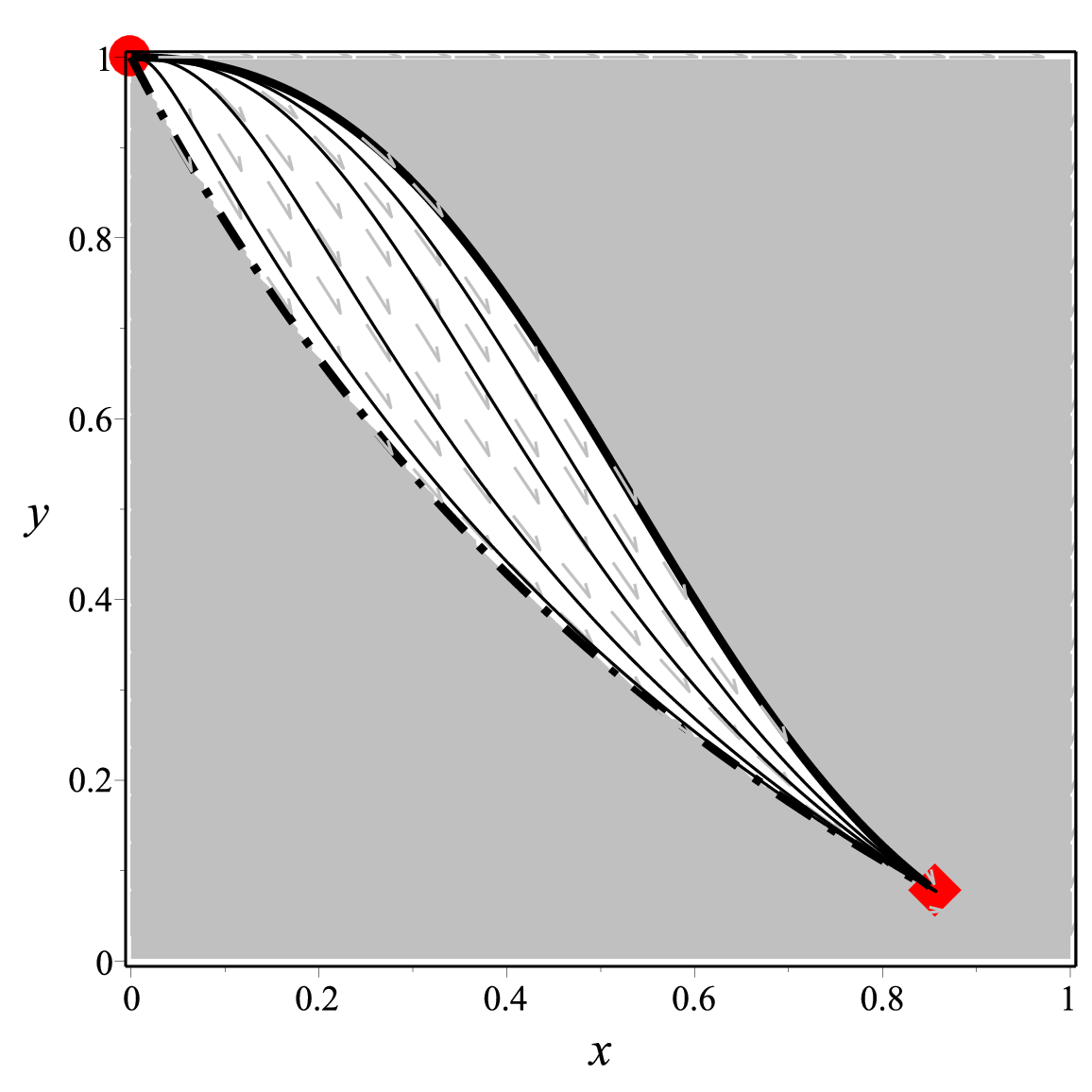}
\caption{Phase portrait of the dynamical system \eqref{dsyst}. The circle encloses the critical point $P_\text{rad}$ in TAB. \ref{tab1} while the solid circle is for the past attractor (source point) $P_\text{bb}$. The solid diamonds mark the location of the future attractors $P_\text{mat}$, $P_\text{dS}$ and $P_\text{K}$, respectively. The black solid curve is the graph of the function $y_\dag=y_\dag(x)$ \eqref{y-dag}, while the dash and dash-dot curves are for the functions $y_\bullet=y_\bullet(x)$ and $y_*=y_*(x)$ in \eqref{curves}, respectively. The left panel shows the whole phase plane, while the middle one represents the physical phase plane $\psi_\text{phys}$ \eqref{psi}, in which the regions where the following conditions: $\Omega_m\geq 0$ and $c^2_s\geq 0$, are not satisfied, have been removed. The right panel shows the region of the physical phase plane $\psi_\text{phys}$, where the stability conditions $c^2_s\geq 0$ and $K_X+2XK_{XX}\geq 0$, as well as the physical requirement that $\Omega_m\geq 0$, are satisfied at once. Notice that the dash curve, which represents the extremum of the energy density of k-essence \eqref{rho-k}, is a separatrix in the phase plane. The fact the the phase plane orbits depart from this curve means that critical points in there represent unstable states.}\label{fig1}
\end{figure*}

The phase plane $\psi$ where to look for the critical points of the dynamical system \eqref{dsyst} is the bounded region $0\leq x\leq 1,$ $0\leq y\leq 1.$ However, if the physical requirements are considered that the squared speed of sound and the dimensionless energy density parameter must be nonnegative quantities, the resulting physical phase space is given by the following region (middle panel of FIG. \ref{fig1}):

\begin{align} \psi_\text{phys}=\{(x,y)|0\leq x\leq 1,\;0\leq y\leq 1,\;\Omega_m\geq 0,\;c^2_s\geq 0\}.\label{psi}\end{align}


\subsection{Critical points}
\label{subsect-cpoints-1}


The critical points of the dynamical system \eqref{dsyst}, as well as their main properties, are summarized in TAB. \ref{tab1}, while the phase portrait of the dynamical system is shown in FIG. \ref{fig1}. We list them here:

\begin{enumerate}

\item Point dominated by matter $P_\text{mat}(0,0)$ $\Rightarrow$ $\Omega_m=1$. It is a local (future) attractor. From equations \eqref{xy-om} and \eqref{xy-hdot} it follows that,

\begin{align} \frac{\dot H}{H^2}&=-\frac{3}{2}(w_m+1)\nonumber\\
&\Rightarrow\;H=\frac{2}{3(w_m+1)(t-t_0)}\nonumber\\
&\Rightarrow\;a\propto(t-t_0)^\frac{2}{3(w_m+1)},\nonumber\end{align} where $t_0$ is an integration constant, while from \eqref{xy-xdot} we have that,

\begin{align} \frac{\dot X}{X}=-6H\;\Rightarrow\;X\propto a^{-6}.\nonumber\end{align} Hence, the k-essence behaves like a stiff fluid. However, this component very quickly decays with the cosmic expansion. Asymptotically ($t\rightarrow\infty$) the matter-dominated static solution is the stable state.

\item Matter-dominated bigbang $P_\text{bb}(0,1)$ $\Rightarrow$ $\Omega_m=1$. In this case,

\begin{align} H=\frac{2}{3(t-t_0)}\;\Rightarrow\;a\propto (t-t_0)^{2/3},\nonumber\end{align} so that, asymptotically into the past ($t\rightarrow t_0$) $H\rightarrow\infty$. This means that the origin of every orbit in the phase plane leading to stable cosmological evolution, is the bigbang singularity.

\item K-essence dominated solution $P_\text{k}(1,0).$ This represents a static universe since, asymptotically, $H\rightarrow 0$. There is not much information we can retrieve from the dynamical systems analysis in this case, but that it is a local future attractor.

\item Radiation-like k-essence $P_\text{rad}(1,1).$ The only useful information we can obtain about this equilibrium configuration is that the k-essence fluid behaves like radiation since $w_\text{eff}=1/3$, that the SSS $c^2_s=1/3$ and that it is a local attractor in the phase plane. Besides, it is obtained also that,

\begin{align} \frac{\dot X}{HX}=-2\;\Rightarrow\;X\propto a^{-2}.\nonumber\end{align} This means that asymptotically,

\begin{align} K(X)=-\alpha X+\beta X^2\propto-\frac{1}{a^2},\nonumber\end{align} or, if take into account the Friedmann equation (first equation in \eqref{eom},) we get that,

\begin{align} 3H^2\approx-\frac{\alpha}{a^2}+\rho_m,\nonumber\end{align} so that the k-essence can be understood as a spatial curvature term.

\item The de Sitter solution $P_\text{dS}(6/7,1/13)$, is a local (future) attractor. It is distinguished by the vanishing sound speed. Condition $x=6/7$ leads to the following:

\begin{align} \frac{\alpha X}{\alpha X+H^2}=\frac{6}{7}\;\Rightarrow\;3H^2=\frac{\alpha}{2}X,\nonumber\end{align} while the condition $y=1/13$ yields,

\begin{align} \frac{\beta H^2}{\beta H^2+\alpha^2}=\frac{1}{13}\;\Rightarrow\;3H^2=\frac{\alpha^2}{4\beta}.\nonumber\end{align} Comparing both results we get that $X=\alpha/2\beta$. This means that the de Sitter attractor $P_\text{dS}$ corresponds to the stable attractor described in \cite{scherrer_prl_2004} for the polynomial model \eqref{pwl-mod}.

\end{enumerate} 

Although at first sight it may seem that the model \eqref{pwl-mod} has a rich structure in the phase plane: five critical points, including an isolated source and three local attractors plus a saddle point (left panel of FIG. \ref{fig1}), the fact is that minimal physical requirements such as $\Omega_m\geq 0$, plus plausible stability conditions such as $c^2_s\geq 0$ and $K_X+2XK_{XX}\geq 0$, significantly reduce the phase space to the narrow region seen in the right panel of FIG. \ref{fig1}. This is the physically meaningful phase plane region where we look for relevant asymptotic solutions. There are only two of them: the source point (matter-dominated bigbang) and the future de Sitter attractor. 

Notice that there are no critical (saddle) points which could be associated with a matter-dominated stage of the cosmic expansion, that is required for the observed amount of cosmic structure to form. The only critical point characterized by matter domination is the past attractor $P_\text{bb}$ that is associated with the initial bigbang singularity. This is contrary to the feature adjudicated to this type of kinetically driven k-essence models that dark matter and dark energy can be explained in a unified way \cite{scherrer_prl_2004}. The fact that, on the basis of a fairly general analysis in the latter reference, it is concluded that the energy density of the k-essence can be written as (equation (22) of \cite{scherrer_prl_2004}) $\rho_\text{eff}=H^2_0+C_0a^{-3}$, does not contradict our results. This only means that the latter behavior can be a very specific solution which can be attained under very specific initial conditions and does not represent a generic result. In contrast, the critical points of the corresponding dynamical system \eqref{dsyst} represent generic behavior.


\begin{figure*}[tbh]\centering
\includegraphics[width=5.5cm]{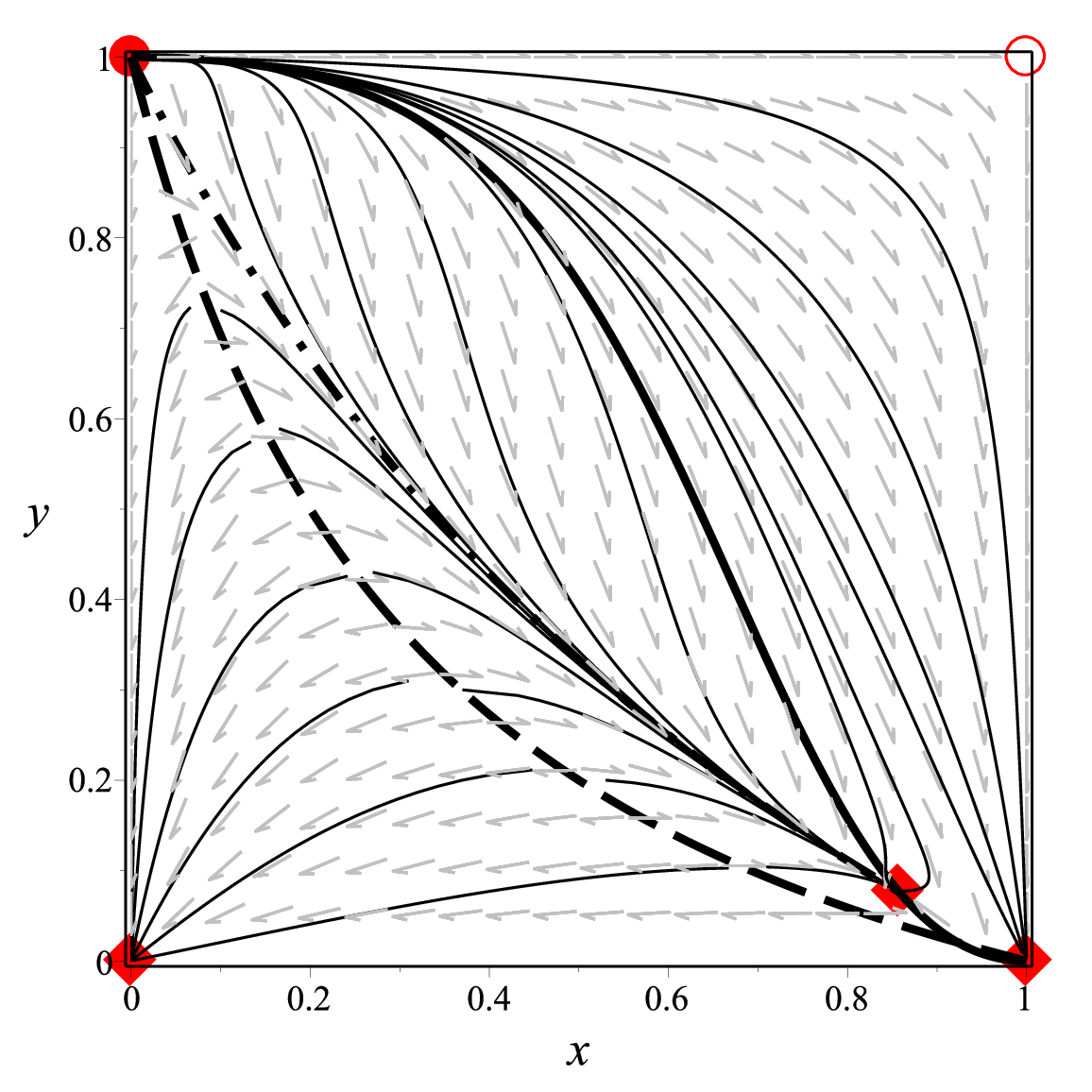}
\includegraphics[width=5.5cm]{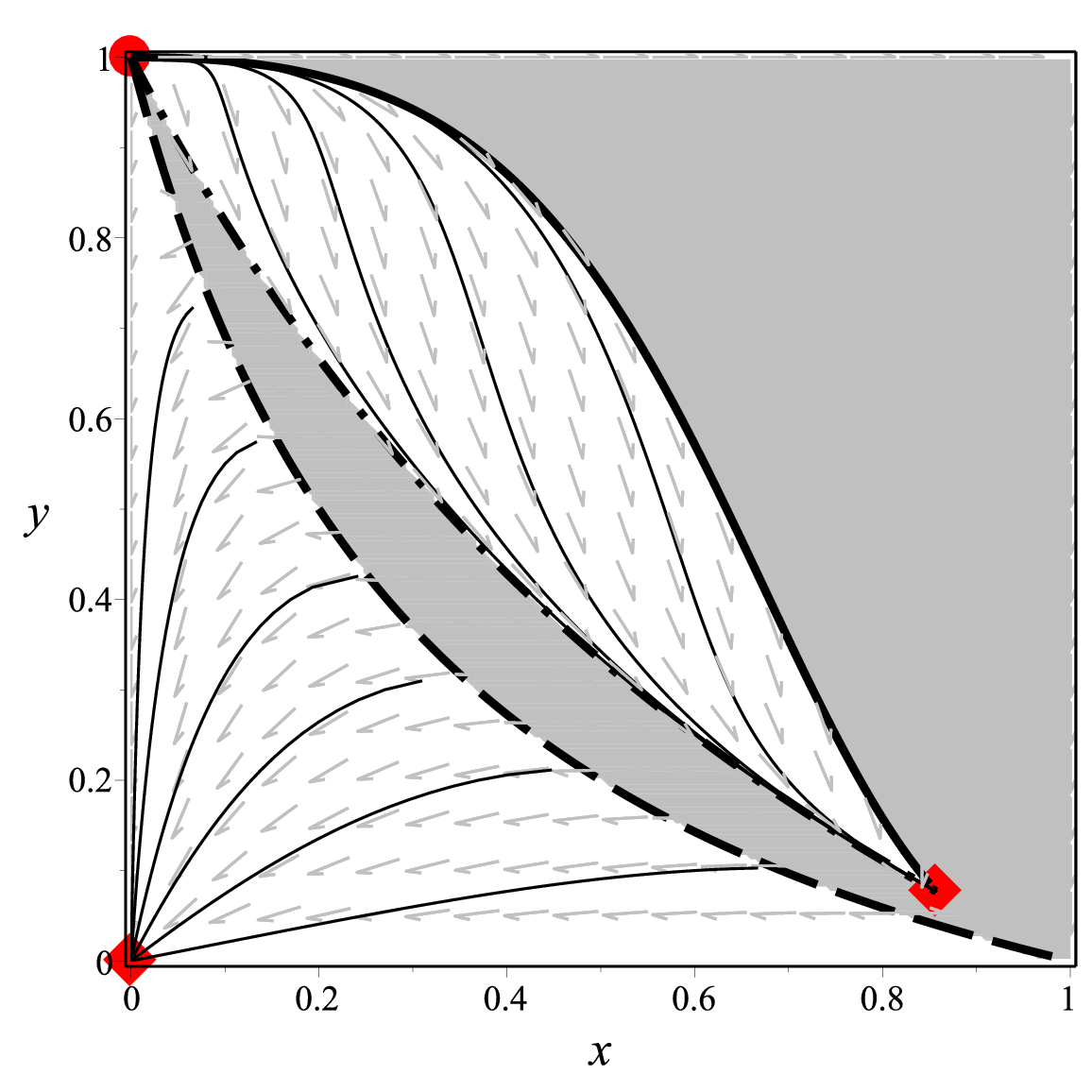}
\includegraphics[width=5.5cm]{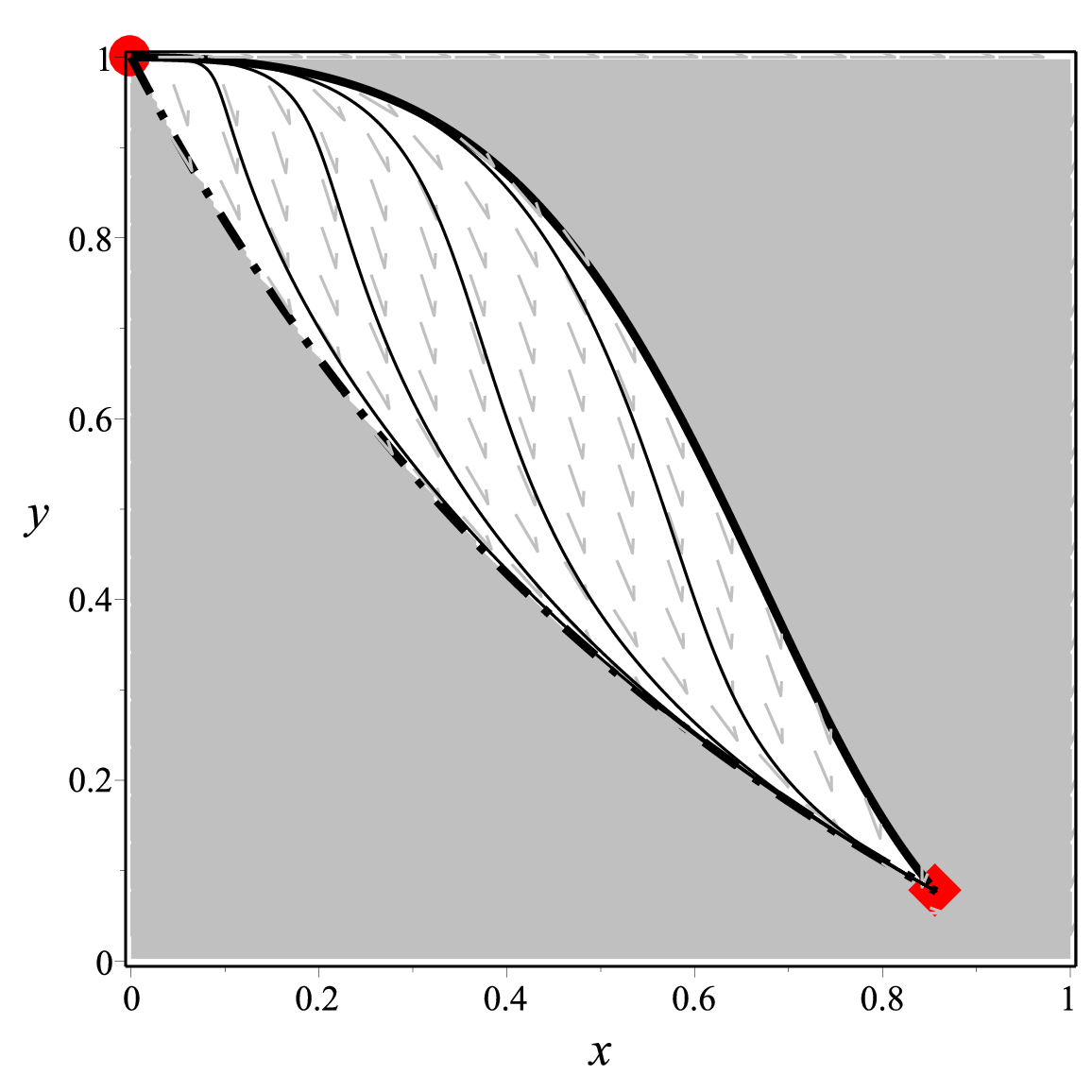}
\caption{Phase portrait of the dynamical system \eqref{asode-2}. The critical points are the same as those of \eqref{dsyst}. The left panel shows the whole phase plane, while the middle one represents the physical phase plane, in which the regions where the following conditions: $\Omega_m\geq 0$ and $c^2_s\geq 0$, are not satisfied, have been removed. The right panel shows the region of the physical phase plane, where the stability conditions $c^2_s\geq 0$ and $K_X+2XK_{XX}\geq 0$, as well as the physical requirement that $\Omega_m\geq 0$, are satisfied at once.}\label{fig2}
\end{figure*}


\section{Model $K=-\alpha\sqrt{X}+\beta X$}
\label{sect-mod-2}


Let us investigate the phase-space dynamics of another model that has been claimed to explain DM and DE in a unified way. In \cite{bose_prd_2009} the k-essence model,

\begin{align} K(\phi,X)=-\alpha\sqrt{X}+\beta X+\lambda-\frac{1}{2}m^2\phi^2,\label{bose}\end{align} where $\alpha\geq 0$, $\beta\geq 0$, $\lambda$ and $m$ are free parameters ($\alpha$ has dimensions of $H$, while $\beta$ is dimensionless,) was proposed as a unifying model of inflation DM and DE. The potential energy domination $V\gg X$ can explain the primordial inflation. At the end of inflation a kinetic-domination stage starts. At this stage one may consider that the potential vanishes $V\sim 0$. The authors of \cite{bose_prd_2009} argue that at this stage the energy density of k-essence goes like (equation (36) of that bibliographic reference) $\rho_\text{eff}\approx 
C_0+C_1 a^{-3}+C_2 a^{-6}$ (the last term very quickly dilutes with cosmic expansion) so that DE and DM can be explained in a unified way, as claimed in \cite{scherrer_prl_2004}. 

Here we shall set $\lambda=m=0,$ so that the last two terms in \eqref{bose} are omitted.\footnote{This choice of parameters does not affect the analysis in \cite{bose_prd_2009} regarding the unified description of DM and DE. Under this choice equation (36) of this reference, which takes place after the end of inflation ($V\rightarrow 0$), reads $$\rho_\text{eff}=\frac{\alpha^2}{4\beta}+\frac{k\alpha}{\beta}\,a^{-3}+\frac{k^2}{\beta}\,a^{-6},$$ where the constant $k$ comes from equation (10) of \cite{bose_prd_2009}: $\sqrt{X}K_X=k a^{-3}$.} The resulting pure k-essence model with $K=K(X)$ belongs to the class discussed in \cite{scherrer_prl_2004}. In this case the effective pressure $p_\text{eff}=K(X)$ is a minimum at $X_0=\alpha^2/4\beta^2.$ At this point $K_X(X_0)=0,$ while $K_{XX}(X_0)=2\beta^3/\alpha^2$ and the effective pressure is negative $K(X_0)=-\alpha^2/4\beta.$ As discussed in \cite{scherrer_prl_2004} this is a stable state, since perturbations around this point $X\rightarrow X_0+\varepsilon$ satisfy the following equation: $\dot\varepsilon=-3H_0\varepsilon,$ where $H_0=\alpha/2\sqrt{3\beta}.$ This point corresponds to a de Sitter attractor solution. 

The curious fact about this model is that if we consider the Friedmann equation alone, since the effective energy density of the k-essence $\rho_\text{eff}=2XK_X-K=\beta X$: $3H^2=\beta X+\rho_m,$ so that it is just a standard scalar field without self-interaction (set $\beta=1).$ Normally this model would not lead to accelerated expansion. However, the Raychaudhuri equation (second equation in \eqref{eom},) as well as the Klein-Gordon type equation \eqref{kg-eom}, differ from those of a standard scalar field without self-interaction. This is why the accelerated expansion solution (the de Sitter attractor) is obtained.

The phase space of this model is very similar to that of model \eqref{pwl-mod}. Let us introduce the following dimensionless and bounded variables of the phase plane (the second variable is the same as in \eqref{xy-var}), 

\begin{align} x=\frac{\alpha\sqrt{X}}{\alpha\sqrt{X}+H^2},\;y=\frac{\beta H^2}{\beta H^2+\alpha^2},\label{xy-var-2}\end{align} so that

\begin{align} \frac{\alpha\sqrt{X}}{H^2}=\frac{x}{1-x},\;\beta H^2=\frac{\alpha^2y}{1-y}.\nonumber\end{align} In terms of these variables we have that,

\begin{align} w_\text{eff}=&\frac{x+y-1}{xy},\;\Omega_\text{eff}=\frac{x^2y}{3(1-x)^2(1-y)},\nonumber\\
\Omega_m=&1-\Omega_\text{eff}.\label{quant}\end{align} The squared sound speed reads,

\begin{align} c^2_s=2\frac{1-x-y-xy}{1-x-y-3xy}.\label{sss-2}\end{align} The following relationships are useful as well:

\begin{align} \frac{\dot H}{H^2}=&-\frac{3}{2}(w_m+1)+\frac{x(1-x-y+w_m xy)}{2(1-x)^2(1-y)},\nonumber\\
\frac{\dot X}{HX}=&-6c^2_s=-12\frac{1-x-y-xy}{1-x-y-3xy}.\label{usef-2}\end{align} 

In this case, the original cosmological equations are traded by the following plane-autonomous dynamical system:

\begin{align} x'=&\frac{1}{2}x(1-x)^3(1-y)\left(\frac{\dot X}{HX}-4\frac{\dot H}{H^2}\right),\nonumber\\
y'=&2y(1-x)^2(1-y)^2\frac{\dot H}{H^2},\label{asode-2}\end{align} where the prime denotes the derivative with respect to the time variable $\tau$, such that $d\tau=d\ln a/(1-x)^2(1-y).$ The physical phase space is the same as in the former model, just recall that the $x$-variable differs a bit from the model \eqref{pwl-mod}. Taking into account this fact, it is easy to find that the critical points of \eqref{asode-2} are exactly the same as those of \eqref{dsyst} for the former model (see TAB. \ref{tab1}). 

For the present model, as for the model \eqref{pwl-mod}, there are no critical points associated with matter domination, but for the bigbang singularity and the matter-dominated attractor, which is not of interest if we take into account the no-ghost condition (left-hand side of \eqref{stab}), required for theoretical consistency \cite{defelice_2012}. A matter-dominated saddle critical point, corresponding to a transient stage of the cosmic expansion, is required to explain the observed amount of cosmic structure.

Once again, the claimed unified explanation of DM and DE does not represent an asymptotic state of the dynamical system \eqref{asode-2}. What is going on then? Although the quite general explanation given in \cite{bose_prd_2009} does actually support the conclusion about the unified description of the dark cosmological sector, it is based on assumptions and simplifications. Each assumption removes a large number of initial conditions that lead to the unified description. Then the state with energy density (equation (36) of the aforementioned bibliographic reference under our choice of constant parameters)

\begin{align} \rho_\text{eff}\approx\frac{\alpha^2}{4\beta}+\frac{k\alpha}{\beta}\,a^{-3}+{\cal O}(a^{-6}),\nonumber\end{align} can be attained under very specific initial conditions and does not represent generic behavior of the dynamical system. We recall that the critical points (equilibrium or singular states) of the dynamical system \eqref{asode-2} represent generic asymptotic behavior. In contrast, any point of the phase plane $\psi_\text{phys}$ defined in \eqref{psi} represents a possible solution of the original set of cosmological equations.


\section{Other models}
\label{sect-other}


There are also interesting models of power-law type. In particular, the model where the effective k-essence pressure is given by

\begin{align} K(X)=\gamma X^n,\label{copel}\end{align} where $\gamma$ and $n$ are free parameters, has been studied in several contexts \cite{chime_mpla_2004, chime_prd_2004, copel_jcap_2024}. For this model we have that $XK_X=nK$ and $X^2K_{XX}=n(n-1)K,$ so that, for instance, the squared speed of light \eqref{sss} and the EOS parameter \eqref{wk}, are constants:

\begin{align} c^2_s=\frac{1}{2n-1}=w_\text{eff}.\label{sss-n}\end{align} The EOM \eqref{eom} read,

\begin{align} 3H^2&=(2n-1)K+\rho_m,\nonumber\\
-2\dot H&=2nK+(w_m+1)\rho_m,\nonumber\\
\dot K&=-\frac{6n}{2n-1}\,HK.\label{eom-n}\end{align} The deceleration parameter is given by

\begin{align} q=-1-\frac{\dot H}{H^2}=&-1+\frac{3}{2}(w_m+1)\nonumber\\
&+\left[\frac{3n}{2n-1}-\frac{3}{2}(w_m+1)\right]\Omega_\text{eff}.\label{q-n}\end{align} The Friedmann constraint following from \eqref{eom-n} can be written as:

\begin{align} \Omega_m+\Omega_\text{eff}=1,\label{fried-n}\end{align} where $\Omega_m$ is the dimensionless energy density of matter and 

\begin{align} \Omega_\text{eff}=\frac{(2n-1)K}{3H^2},\label{ok-n}\end{align} is the dimensionless energy density of the k-essence. The following autonomous ordinary differential equation (ODE) on $\Omega_\text{eff}$ is easily obtained:

\begin{align} \Omega'_\text{eff}=\lambda\Omega_\text{eff}(1-\Omega_\text{eff}),\label{ode-n}\end{align} where the prime denotes derivative with respect to the variable $N=\ln a$ and we have introduced the constant parameter,

\begin{align} \lambda=3w_m-\frac{3}{2n-1}.\label{lambda-n}\end{align} 

There are two critical points of the ODE \eqref{ode-n}: 1) the matter-dominated solution $\Omega_\text{eff}=0$, which is stable if $\lambda<0$ is a negative quantity and unstable otherwise, and 2) the k-essence-dominated solution $\Omega_\text{eff}=1$, which is a stable solution if $\lambda\geq 0$ and unstable otherwise. For the matter-dominated solution the deceleration parameter is given by $q=(3w_m+1)/2,$ while for the k-essence-dominated one it is $q=(n+1)/(2n-1)$. This means that none of the above solutions yields accelerated expansion, and so are devoid of cosmological interest as models of dark energy. In addition, for the background dust fluid with $w_m=0,$ the matter-dominated solution is the future attractor, while the k-essence-dominated solution is the past attractor in the phase segment $0\leq\Omega_\text{eff}\leq 1$. Only consideration of the nonvanishing self-interaction potential, as in \cite{copel_jcap_2024}: $K(\phi,X)=\lambda X^n-V(\phi)$, can yield interesting dynamics with the tracking behavior being a stable point of the resulting dynamical system. However, k-essence models with the inclusion of a non-vanishing self-interaction potential lose their original motivation as models that yield accelerated expansion on the basis of the (non-canonical) kinetic energy term alone.



\section{Discussion and conclusion}
\label{sect-discuss}


Generalizations of the kinetic k-essence model \eqref{pwl-mod} have been investigated in the bibliography. For example, in \cite{yang_2011} the asymptotic dynamics of the model with effective pressure. 

\begin{align} K(\phi,X)=V(\phi)(-X+X^2),\label{yang}\end{align} was investigated, where the self-interaction potential was taken to be $V\propto\phi^{-2}$. In this case the quantity $\lambda\equiv V_\phi/V^{3/2}$ is a constant. The polynomial model \eqref{pwl-mod} corresponds to the particular case when in \eqref{yang} $V=$const., so that $\lambda=0$. The critical points of the corresponding dynamical system are summarized in TAB. I of \cite{yang_2011}. There are only three critical points for $\lambda=0.$ Two of these points, $P_2(1,0)$ and $P_3(-1,0)$ correspond to just one critical point in our TAB. \ref{tab1}: $P_\text{bb}$, due to the definition of the phase space variables in \cite{yang_2011}: $\bar x=\dot\phi=\pm\sqrt{2X}$ and $\bar y=\sqrt{V}/\sqrt{3}H$. The third point in TAB. I of \cite{yang_2011} $P_1(0,0)$ corresponds to the point $P_\text{mat}$ in TAB. \ref{tab1} of the present paper. This means that several critical points of the dynamical system corresponding to the model \eqref{pwl-mod}, are lacking in the mentioned bibliographic reference. Although only one of these lacking points: $P_\text{dS}$, belongs in the physically meaningful region, it is the future attractor which can explain the accelerated pace of the present stage of the cosmic expansion. The explanation of why the most distinctive point of this model, which is moreover the future attractor \cite{scherrer_prl_2004}, was not found in \cite{yang_2011}, may be based on their choice of the phase space variables which, we suppose, do not cover the whole physically meaningful phase space. 

In \cite{fang_2014} the more general model \eqref{yang} was studied where $V=V(\phi)$ can be any function. The authors of that paper chose the cosmological parameters $\Omega_\text{eff},$ $\gamma_K=w_\text{eff}+1$ and $\lambda\equiv V_\phi/V^{3/2}$ as variables of the three-dimensional phase space. The model studied in the present paper corresponds to the particular case of the one studied in \cite{fang_2014} when $V=$ const. $\Rightarrow$ $\lambda=0$. Only one physically significant critical point was found in that bibliographic reference, which corresponds to one of the critical points found in our present study. It was, precisely, the de Sitter solution $P_\text{dS}.$ The remaining points are missing in \cite{fang_2014} for the case where $\lambda=0$. 

In the bibliographic reference \cite{chakra_2019} the model:

\begin{align} K(\phi,X)=-\alpha(\phi)X+\beta(\phi)X^2-V(\phi),\label{chakra}\end{align} was investigated. In the case when $V=0$ and the parameters 

\begin{align} \lambda=\frac{\alpha_\phi}{\alpha}\frac{\dot\phi}{H},\;\delta=\frac{\beta_\phi}{\beta}\frac{\dot\phi}{H},\nonumber\end{align} are constants, the obtained ASODE is a two-dimensional dynamical system. In this case, which is a bit more general than the model studied in the present paper, only three kinetic energy dominated ($\Omega_\text{eff}=1$) critical points, are found. One of these corresponds to the de Sitter attractor $P_\text{dS}$. The bigbang source point $P_\text{bb}$ is missing again. In none of the previous dynamical systems studies of purely kinetic k-essence, the stability conditions \eqref{stab} were taken into consideration. This is why in these studies there are declared more critical points than the ones having physical meaning.


On the basis of our study we may conclude that polynomial and power-law models, such as \eqref{pwl-mod}, \eqref{bose} and \eqref{copel}, are not adequate to reproduce the cosmic evolution of our universe due either to the lack of a matter-dominated saddle critical point, as in models \eqref{pwl-mod} and \eqref{bose} or to the inability to produce accelerated expansion, as in model \eqref{copel}. In the former case, a transient stage of matter domination is required in order for the amount of cosmic structure found in astrophysical observations to form. Hence, one could expect the existence of a saddle critical point characterized by matter dominance $\Omega_m=1$, in the physically meaningful phase space, which is also required for a unified description of DM and DE. In the present case only a past attractor (the bigbang singularity) and a future de Sitter attractor are found (right panel of FIG. \ref{fig1}). 

Our conclusion adds to other conclusions pointing in the same direction: k-essence models are not attractive models to explain the amount of existing cosmological observational data, unless one adds other elements already present in other competing models. In addition, observations, in particular local (model-independent) determinations of the present value of the Hubble constant $H_0$, disfavor models with an equation of state $w(z)>-1$. This includes the quintessence and k-essence models \cite{colgain}.


\section*{Acknowledgments}


The authors thank Jorge Cervantes and Eoin O Colgain for useful comments. We also acknowledge FORDECYT-PRONACES-CONACYT for support of the present research under grants CF-MG-2558591 and CF-140630-UNAM-UMSNH.





\begin{thebibliography}{99}


\bibitem{picon_damour_plb_1999} C. Armendariz-Picon, T. Damour, V.F. Mukhanov, Phys. Lett. B {\bf 458} (1999) 209-218 [e-Print: hep-th/9904075]

\bibitem{garriga_plb_1999} J. Garriga, V.F. Mukhanov, Phys. Lett. B {\bf 458} (1999) 219-225 [e-Print: hep-th/9904176]

\bibitem{chiba_prd_2000} T. Chiba, T. Okabe and M. Yamaguchi, Phys. Rev. D {\bf 62} (2000) 023511 [e-Print: astro-ph/9912463]

\bibitem{picon_mukhanov_prl_2000} C. Armendariz-Picon, V.F. Mukhanov, P.J. Steinhardt, Phys. Rev. Lett. {\bf 85} (2000) 4438-4441 [e-Print: astro-ph/0004134]

\bibitem{picon_mukhanov_prd_2001} C. Armendariz-Picon, V.F. Mukhanov, P.J. Steinhardt, Phys. Rev. D {\bf 63} (2001) 103510 [e-Print: astro-ph/0006373]

\bibitem{chime_mpla_2004} L.P. Chimento and A. Feinstein, Mod. Phys. Lett. A {\bf 19} (2004) 761-768 [e-Print: astro-ph/0305007]

\bibitem{chime_prd_2004} L.P. Chimento, Phys. Rev. D {\bf 69} (2004) 123517 [e-Print: astro-ph/0311613]

\bibitem{bose_prd_2009} N. Bose and A.S. Majumdar, Phys. Rev. D {\bf 79} (2009) 103517 [e-Print: 0812.4131]

\bibitem{scherrer_prl_2004} R.J. Scherrer, Phys. Rev. Lett. {\bf 93} (2004) 011301 [e-Print: astro-ph/0402316]

\bibitem{defelice_2012} A. De Felice and S. Tsujikawa, JCAP {\bf 02} (2012) 007 [e-Print: 1110.3878]

\bibitem{fang_2014} W. Fang, H. Tu, Y. Li, J. Huang and C. Shu, Phys. Rev. D {\bf 89} (2014) 123514 [e-Print: 1406.0128]

\bibitem{erickson_2002} J.K. Erickson, R.R. Caldwell, P.J. Steinhardt, C. Armendariz-Picon, V.F. Mukhanov, Phys. Rev. Lett. {\bf 88} (2002) 121301 [e-Print: astro-ph/0112438]

\bibitem{dedeo_2004} S. DeDeo, R.R. Caldwell, P.J. Steinhardt, Phys. Rev. D {\bf 67} (2003) 103509; Phys. Rev. D {\bf 69} (2004) 129902 (erratum) [e-Print: astro-ph/0301284]

\bibitem{yang_2011} R.J. Yang and X.T. Gao, Class. Quant. Grav. {\bf 28} (2011) 065012 [e-Print: 1006.4986]

\bibitem{chakra_2019} A. Chakraborty, A. Ghosh and N. Banerjee, Phys. Rev. D {\bf 99} (2019) 103513 [e-Print: 1904.10149]

\bibitem{cervantes_prd_2013} J. De-Santiago, J.L. Cervantes-Cota, D. Wands, Phys. Rev. D {\bf 87} (2013) 023502 [e-Print: 1204.3631]

\bibitem{wands_prd_1998} E.J. Copeland, A.R. Liddle and D. Wands, Phys. Rev. D {\bf 57} (1998) 4686-4690 [e-Print: gr-qc/9711068]

\bibitem{copeland_rev_2006} E.J. Copeland, M. Sami and S. Tsujikawa, Int. J. Mod. Phys. D {\bf 15} (2006) 1753-1936 [e-Print: hep-th/0603057]

\bibitem{quiros_ejp_2015} R. García-Salcedo, T. Gonzalez, F.A. Horta-Rangel, I. Quiros and D. Sanchez-Guzmán, Eur. J. Phys. {\bf 36} (2015) 025008 [e-Print: 1501.04851]

\bibitem{quiros_rev_2019} I. Quiros, Int. J. Mod. Phys. D {\bf 28} (2019) 1930012 [e-Print: 1901.08690]

\bibitem{copel_jcap_2024} E.J. Copeland, A. Moss, S. Sevillano Muñoz and J.M.M. White, JCAP {\bf 05} (2024) 078 [e-Print: 2309.15295]

\bibitem{colgain} B.H. Lee, W. Lee, E.Ó. Colgáin, M.M. Sheikh-Jabbari, S. Thakur, JCAP {\bf 04} (2022) 004 [e-Print: 2202.03906]



\end{thebibliography}
\end{document}